\documentclass[12pt]{article}
\usepackage{graphicx}

\newcommand\pubnumber{SNSN-323-63}
\newcommand\pubdate{\today}

\def\cycl{Cyclotron Institute, Texas A\&M University\\
College Station, TX, 77843-3366 U.S.A.}
\def\support{\footnote{This material is based upon work
    supported by the U.S. Department of Energy, Office of Science, Office of Nuclear Physics, under Award Number DE-FG03-93ER40773, 
    and by the Welch Foundation under Grant No.~A-1397.}}

\def\Title#1{\begin{center} {\Large #1 } \end{center}}
\def\Author#1{\begin{center}{ \sc #1} \end{center}}
\def\Address#1{\begin{center}{ \it #1} \end{center}}

\newcommand\pubblock{\rightline{\begin{tabular}{l} \pubnumber\\
         \pubdate  \end{tabular}}}
\newenvironment{Abstract}{\begin{quotation}  }{\end{quotation}}
\newenvironment{Presented}{\begin{quotation} \begin{center} 
             PRESENTED AT\end{center}\bigskip 
      \begin{center}\begin{large}}{\end{large}\end{center} \end{quotation}}

\def\beq{\begin{equation}}
\def\eeq#1{\label{#1}\end{equation}}
\def\eeqn{\end{equation}}

\def\beqa{\begin{eqnarray}}
\def\eeqa#1{\label{#1}\end{eqnarray}}
\def\eeqan{\end{eqnarray}}

\let\bar=\overbar

\def\Dslash{\not{\hbox{\kern-4pt $D$}}}
\def\dslash{\not{\hbox{\kern-2pt $\del$}}}

\def\ee{e^+e^-}

\def\GF{G_F}

\def\msb{{\bar{\ssstyle M \kern -1pt S}}}

\begin{document}

\def\QEC{Q_{\mbox{\tiny EC}}}
\def\GV{G_{\mbox{\tiny V}}}
\def\Vud{V_{\mbox{\scriptsize ud}}}
\def\Vus{V_{\mbox{\scriptsize us}}}
\def\Vub{V_{\mbox{\scriptsize ub}}}
\def\F{{\cal F}}
\def\DRV{\Delta_{\mbox{\tiny R}}^{\mbox{\tiny V}}}
\def\GF{G_{\mbox{\tiny F}}}
\def\be {\begin{equation}}
\def\ee {\end{equation}}
\def\bea {\begin{eqnarray}}
\def\eea {\end{eqnarray}}

\begin{titlepage}
\pubblock

\vfill
\Title{Nuclear beta decays and CKM unitarity}
\vfill
\Author{ J.C. Hardy\support and I.S. Towner$^1$}
\Address{\cycl}
\vfill
\begin{Abstract}
Nuclear $\beta$ decays between ($J^{\pi},T$) = ($0^+,1$) analog states yield the best value for the $\Vud$ element of the
Cabibbo-Kobayashi-Maskawa matrix. Current world data establish the corrected $\F t$ values of 14 separate superallowed transitions
to a precision of order 0.1\% or better. The validity of the small theoretical correction terms is confirmed by excellent
consistency among the 14 $\F t$ values and by recent measurements that compare pairs of mirror superallowed transitions. With
consistency established, the results now yield $|\Vud|$ = 0.97420(21). This value is consistent with the considerably less
precise results obtained from $\beta$ decays of the neutron, the pion and $T$=1/2 mirror nuclei, which are hampered by experimental
challenges.

\end{Abstract}
\vfill
\begin{Presented}
CIPANP2018 \\
Thirteenth Conference on the Intersections of Particle and Nuclear Physics  \\ [2mm]
Palm Springs CA, U.S.A., May 29 -- June 3, 2018
\end{Presented}
\vfill
\end{titlepage}
\def\thefootnote{\fnsymbol{footnote}}
\setcounter{footnote}{0}

\section{Superallowed nuclear beta decay}

Beta decay between nuclear analog states of spin-parity, $J^{\pi} = 0^+$, and isospin, $T = 1$, has a unique simplicity: It is a pure
vector transition and is nearly independent of the nuclear structure of the parent and daughter states.  Such transitions are called
``superallowed."  Their measured strength -- expressed as an ``$ft$ value" -- can be related directly to the vector coupling constant
for semi-leptonic decays, $\GV$, with the intervention of only a few small ($\sim$1\%) calculated terms to account for radiative and
nuclear-structure-dependent effects. Once $\GV$ has been determined in this way, it is only another short step to obtain a value for
$\Vud$, the up-down element of the Cabibbo-Kobayashi-Maskawa (CKM) quark-mixing matrix, and then use it to test CKM unitarity via the
sum of squares of the top-row elements, $\Vud^2$+$\Vus^2$+$\Vub^2$.  Any deviation of this sum from unity would signal the presence of
physics beyond the Standard Model.

The $ft$ value of any $\beta$ transition is simply the product of the phase-space factor, $f$, and the partial half-life of the transition,
$t$.  It depends on three measured quantities: the total transition energy, $\QEC$, the half-life, $t_{1/2}$, of the parent state, and the
branching ratio, $R$, for the particular transition of interest. The $\QEC$ value is required to determine $f$, while the half-life and
branching ratio combine to yield the partial half-life.

In dealing with superallowed decays, it is convenient to combine some of the small correction terms with the measured $ft$-value and
define a corrected $\F t$-value. Thus, we write \cite{HT15}
\be
\F t \equiv ft (1 + \delta_{\mbox{\tiny R}}^{\prime}) (1 + \delta_{\mbox{\tiny NS}} - \delta_{\mbox{\tiny C}} ) = \frac{K}{2 \GV^2 
(1 + \DRV )}~,
\label{Ftconst}
\ee
where $K = 8120.2776(9) \times 10^{-10}$ GeV$^{-4}$s, $\delta_{\mbox{\tiny C}}$ is the isospin-symmetry-breaking correction and $\DRV$
is the transition-independent part of the radiative correction. The terms $\delta_{\mbox{\tiny R}}^{\prime}$ and $\delta_{\mbox{\tiny NS}}$
constitute the transition-dependent part of the radiative correction, the former being a function only of the electron's energy and the $Z$
of the daughter nucleus, while the latter, like $\delta_{\mbox{\tiny C}}$, depends in its evaluation on nuclear structure. From this equation,
it can be seen that a measurement of any one superallowed transition establishes a value for $\GV$. The measurement of several tests the
Conserved Vector Current (CVC) hypothesis that $\GV$ is not renormalized in the nuclear medium. If indeed $\GV$ is constant -- i.e. all the
$\F t$-values are the same -- then an average value for $\GV$ can be determined and $\Vud$ obtained from the relation $\Vud = \GV/\GF$, where
$\GF$ is the well known \cite{PDG18,Ti13} weak-interaction constant for purely leptonic muon decay.

It is important to note that if, instead, the $\F t$ values were to show a significantly non-statistical inconsistency, 
one to the other, then the remaining steps could not be taken since inconsistency would demonstrate that the correction
terms were not correct or, less likely, that CVC had been violated.  Without consistency, there is no coupling
``constant" and there can be no justification for extracting a value for $\Vud$.

\begin{figure}[t]
\centering
\includegraphics[width=13.6 cm]{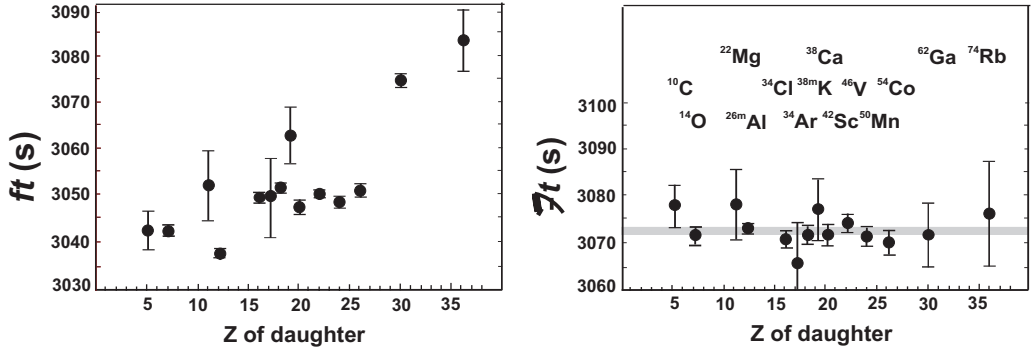}
  \caption{Results from the 2015 survey \protect\cite{HT15} updated through 2017: uncorrected $ft$ values for the 14 best
known superallowed decays on the left; the same results but incorporating the $\delta_{\mbox{\tiny R}}^{\prime}$,
$\delta_{\mbox{\tiny C}}$ and $\delta_{\mbox{\tiny NS}}$ correction terms on the right.  The grey band in the right panel is the average $\F$$t$
value and its uncertainty.}
\label{fig1}
\end{figure}

Early in 2015, we published \cite{HT15} the most recent critical survey of all half-life, decay-energy and branching-ratio measurements related to 20
superallowed $0^+$$\rightarrow 0^+$ $\beta$ decays.  Included were 222 individual measurements of comparable  precision obtained from 177
published references.  Since that time, through 2017, there have been more than a dozen new published measurements of relevance, the most
consequential having impacted the results for $^{10}$C \cite{Du16}, $^{14}$O \cite{Va15, Vo15}, $^{38}$Ca \cite{Bl15} and $^{42}$Sc
\cite{Er17}.  Incorporating these new measurements, we have updated our survey results, with the outcome shown in Fig.~\ref{fig1} for the 14
transitions known with a precision of order 0.1\% or better.  Obviously the $\F$$t$ values are all consistent with one another over the full
measured range from $Z$=5 to $Z$=36.

\begin{figure}[t]
\centering
\includegraphics[width=8 cm]{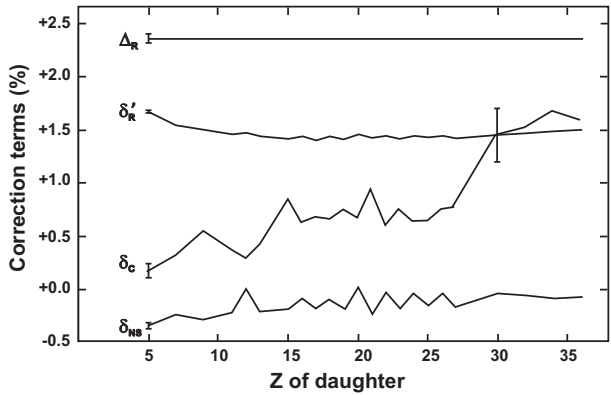}
  \caption{The four calculated correction terms that appear in Eq.\,(\ref{Ftconst}) plotted as a function of the $Z$ of the daughter. The lines
serve only to guide the eye.  A few sample (theoretical) uncertainties are shown.  It should be noted that the magnitude of the uncertainty shown
for $\delta_{\mbox{\tiny C}}$ at $Z$=5 remains qualitatively
unchanged up to $Z$=27, after which it rises to values indicated by the uncertainty shown for $Z$=30.}
\label{fig2}
\end{figure}

Before drawing conclusions, it is helpful to examine Fig.\,\ref{fig2}, which illustrates the transition-to-transition behavior of the
correction terms that appear in Eq.\,(\ref{Ftconst}). Of the three terms that contribute to the $\F t$ values, only $\delta_{\mbox{\tiny C}}$
and $\delta_{\mbox{\tiny NS}}$ show pronounced changes as a function of $Z$, so it is these terms that are principally responsible for
replacing the scatter in $ft$ values with the consistency of the $\F t$ values. As described in Ref.\,\cite{HT15}, both terms were derived
from the best available shell-model wave functions, which in each case had been based on a wide range of spectroscopic data for nuclei in the
same mass region.  The calculations for $\delta_{\mbox{\tiny C}}$ were further tuned to reproduce the measured binding energies, charge radii and
coefficients of the isobaric multiplet mass equation for the specific states involved.  This means that the origins of these correction terms
are completely independent of the superallowed decay data.

Thus the consistency of the $\F t$ values in Fig.\,\ref{fig1}  not only confirms the CVC expectation of a constant value for $\GV$ but also
validates the nuclear-structure-dependent radiative and isospin-symmetry-breaking corrections, which converted the measured $ft$ values into
corrected $\F t$ values. 

A further confirmation of the nuclear-structure-dependent corrections is now afforded by the precise measurements of mirror pairs of
$0^+$$\rightarrow 0^+$ superallowed transitions.  It turns out that the ratio of their $ft$ values is a particularly sensitive test of
($\delta_{\mbox{\tiny NS}} - \delta_{\mbox{\tiny C}}$) \cite{Pa14}.  The first mirror pair to have both members precisely characterized was
$^{38}$Ca $\rightarrow$$^{38m}$K and $^{38m}$K $\rightarrow$$^{38}$Ar \cite{Pa14,Pa15}.  It agreed well with the corrections terms used in the
survey illustrated in Fig.\,\ref{fig1} and disfavored an alternative approach.  Two more mirror pairs, with $A$=26 and $A$=34, have now been
completed and their results will soon be published; preliminarily both arrive at the same conclusion.
 
Finally then, with a mutually consistent set of $\F$$t$ values, one is then justified in proceeding to determine the value of $\GV$ and, 
from it, $\Vud$.  The result we obtain from our updated survey is
\bea
|\Vud| = 0.97420(21)~~~~~~~~~~~~~~~~~~~~~~~~~~~~~~~  \mbox{[nuclear superallowed]}. \nonumber 
\eea

\section{Other methods for determining $\Vud$}

Neutron $\beta$ decay is the simplest $\beta$ decay to involve both the vector and axial-vector weak interactions.  It is
an attractive option for determining $\Vud$ since its analysis does not require the application of corrections for
isospin-symmetry-breaking, $\delta_{\mbox{\tiny C}}$, or for nuclear-structure-dependent radiative effects,
$\delta_{\mbox{\tiny NS}}$.  However, it has the distinct disadvantage that it requires a difficult correlation measurement
in order to separate the vector-current contribution to its decay from the axial-vector one.  Not only that, but neutrons are
inherently more difficult to handle and contain than nuclei.

Since the $\QEC$ value and the branching ratio for neutron $\beta$ decay are very well known, the crucial measurements required
to determine $\Vud$ are its mean-life and a decay correlation -- usually selected to be the $\beta$ asymmetry from the decay of
polarized neutrons.  Currently, world data \cite{PDG18,Se18,Br18,Pa18} for both these quantities are not statistically consistent among themselves, the normalized
chi-squared ($\chi^2/N$) being 4.3 for both the mean-life average and the asymmetry average.  More alarming still is the fact
that the mean-life results from two different measurement techniques appear to be systematically different from one another.  
While the overall average mean-life is 879.8(8)\,s, the average of only those measurements in which the decay products were
recorded from a beam of neutrons is 888.1(20)s; while it is 879.4(7)s when neutrons are confined in a ``bottle" and the survivors
are counted a known time later.  It is difficult to know how to deal with such conflicts so we employ two different methods.  
With the first, we follow exactly the same procedures as we do for the superallowed decays, averaging all world data for each
parameter and increasing its uncertainty by the square root of the normalized chi-squared.  For the second we simply assign a
range to the mean-life, which encompasses both the conflicting sets of results.  The results for $\Vud$ are
\bea 
|\Vud| = 0.9754(13)~~~~~~~~~~~~~~~~~~~~~~~~~~~~~~~  \mbox{[neutron average]},
\nonumber \\
0.9700 \leq |\Vud| \leq 0.9760~~~~~~~~~~~~~~~~~~~~~~~~~~~~~~~  \mbox{[neutron range]}. \nonumber 
\eea

Neutron $\beta$ decay is just a special case of decay between $T=1/2$ mirror nuclei.  Like neutron decay, these nuclear mirror
decays are mixed vector and axial-vector decays; so, in addition to $\QEC$ values, half-lives and branching ratios, they also
require a $\beta$-asymmetry measurement.  Of course, unlike the neutron, these decays as well require the corrections
$\delta_{\mbox{\tiny C}}$ and $\delta_{\mbox{\tiny NS}}$ for small nuclear-structure-dependent effects.  There are four mirror
decays, $^{19}$Ne, $^{21}$Na, $^{35}$Ar and $^{37}$K, for which sufficient data are known.  The relevant world data were
last surveyed extensively a decade ago \cite{Se08,Na09}, but there have been several new experimental results since then.  A
recent published update of the world average \cite{Fe18}, which references all new measurements published to date, yields the
following value of $|\Vud|$ as obtained from the $T=1/2$ mirror decays.
\bea 
|\Vud| = 0.9727(14)~~~~~~~~~~~~~~~~~~~~~~~~~~~~~~~  \mbox{[mirror nuclei]}. \nonumber 
\eea

Finally, the rare pion beta decay, $\pi^+ \rightarrow \pi^0 e^+ \nu_e$, which has a branching ratio of $\sim$$10^{-8}$, is one
of the most basic semi-leptonic electroweak processes.  It is a pure vector transition between two spin-zero members of an isospin
triplet and is therefore analogous to the superallowed $0^+$$\rightarrow$$0^+$ decays.  In principle, it can yield a value of $\Vud$
unaffected by nuclear-structure uncertainties.  In practice, the branching ratio is very small and has proved difficult to measure
with sufficient precision.  The most recent, and by far the most precise, measurement of the branching ratio is by the PIBETA group
\cite{Po04}.  This leads to the result \cite{Bl13}
\bea 
|\Vud| = 0.9749(26)~~~~~~~~~~~~~~~~~~~~~~~~~~~~~~~  \mbox{[pion]}. \nonumber
\eea

\begin{figure}[t]
\centering
\includegraphics[width=11 cm]{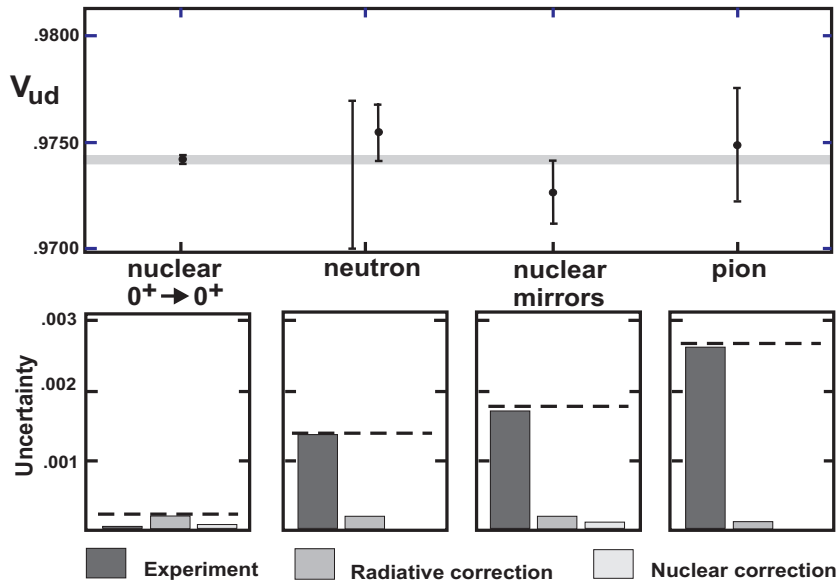}
  \caption{The five values of $|\Vud|$ given in the text are shown in the top panel, the grey band being the average value.  The
four panels at the bottom show the error budgets for the corresponding results with points and error bars at the top.}
\label{fig3}
\end{figure}

\section{Recommended value for $\Vud$}

The five results we have quoted for $|\Vud|$ are plotted in Fig.~\ref{fig3}.  Obviously they are consistent with one another but,
because the nuclear superallowed value has an uncertainty 6 to 12 times smaller than the other results, it dominates the
average.  Furthermore, the more precise of the two neutron results can hardly be considered definitive since it ignores a serious
systematic uncertainty in the data.  Consequently we recommend using the nuclear superallowed result as the best value for $|\Vud|$: i.e.
\be
|\Vud| = 0.97420(21) .
\label{Vud}
\ee

\section{Potential for improvement}

The uncertainty budgets plotted in the bottom panels of Fig.~\ref{fig3} clearly demonstrate that only in the case of the nuclear
$0^+$$\rightarrow$$0^+$ decays is experiment not the predominant source of uncertainty. In fact, by far the most important contribution to the
uncertainty in our recommended value of $\Vud$ is from radiative corrections, principally from $\DRV$, the transition-independent part of the
radiative correction, as can be seen from Fig.\,\ref{fig2}.  Furthermore, the size of the $\DRV$ contribution is approximately the same for all
measurement methods, leaving us to conclude that no major improvement in the value of $|\Vud|$ can be achieved in future by any method without
there being improved calculations of $\DRV$. Such improvements may be on the horizon however \cite{Go18}.

If the uncertainty in $\DRV$ were indeed to be reduced, the impact would be immediate: If it were cut in half, for example, the $|\Vud|$ uncertainty
would be reduced by 30\%. Undoubtedly such an outcome would, in turn, motivate renewed efforts to reduce the experimental uncertainties on the
$ft$ values for the $0^+$$\rightarrow 0^+$ decays. Untimately, though, this method will be limited by the uncertainties associated with the
nuclear-structure-dependent correction terms; and realistically these will be very difficult to improve.

In the long run, without the need for structure corrections, neutron decay (and pion decay) could push the $|\Vud|$ uncertainty still lower but, as
Fig.\,\ref{fig3} makes clear, there is a long way to go in improving the experimental precision of these techniques.  It is hard to predict how far
in the future it will be before such complex experiments can challenge the precision achieved in the more straightforward $0^+$$\rightarrow 0^+$ measurements.

\section{CKM unitarity test} 
  
The standard model does not prescribe the individual elements of the CKM matrix -- they must be determined experimently -- but absolutely fundamental
to the model is the requirement that the matrix be unitary.  To date, the most demanding test of CKM unitarity comes from the sum of squares of
the top-row elements, $|\Vud|^2 + |\Vus|^2 + |\Vub|^2$, which should equal exactly one.  The $|\Vud|$ element is by far the largest of the three. 

The second largest element, $|\Vus|$, can be obtained from a variety of decays but currently two classes of decay predominate.  First, $|\Vus|$ itself
is extracted from semileptonic kaon decays ($K \rightarrow \pi \ell \nu_{\ell}$), while $|\Vus|/|\Vud|$ is obtained from the ratio of the pure leptonic decay
of the kaon ($K^{\pm} \rightarrow \mu^{\pm} \nu$) to to that of the pion ($\pi^{\pm} \rightarrow \mu^{\pm} \nu$).  Both require lattice QCD calculations of
relevant form factors in their analysis. In the past, the results for $|\Vus|$ and $|\Vus|/|\Vud|$ have formed a consistent set with the result for $|\Vud|$.
However, as the quoted uncertainties on the lattice calculations have been reduced, some tension has appeared.  The current Particle Data Group assessment
\cite{Ce18} quotes $|\Vus|$ = 0.2231(8) from the semi-leptonic decays and $|\Vus|$ = 0.2253(7) from the ratio of pure leptonic decays.  If we take the
weighted average of these two values with its uncertainty scaled to account for the large chi-square, we obtain $|\Vus|$ = 0.2243(11).

Finally, incorporating the PDG value, 0.0039(4), for $|\Vub|$, we find the unitary sum to be
\bea 
|\Vud|^2+|\Vus|^2+|\Vub|^2 = 0.99939(64),
\eea

\noindent which confirms unitarity to within $\pm$0.06\%.

\end{document}